\pdfoutput=1 
\documentclass[twocolumn,notitlepage]{article}  
\usepackage{graphicx}
\usepackage{txfonts}

\usepackage{natbib}
\bibpunct{(}{)}{;}{a}{}{,} 
\begin{document}

\title{Proposed high order harmonic interferometer\\
	for aperture synthesis radio telescope:\\
	Theory and computer simulation}

\author{Nailong Wu\\
	Independent Researcher\\
	Toronto, Canada. e-mail: dam16642@yahoo.com}

\twocolumn[
	\begin{@twocolumnfalse}
		\maketitle
		\begin{abstract}

\noindent
A new type of interferometer, called High Order Harmonic Interferometer (HOHI), 
was proposed by \citet{wu1996} for imaging by aperture synthesis radio telescope. 
Its  feasibility was proven by theoretical analysis. Before putting HOHI in 
practical use, computer simulation is a necessary intermediate stage.
In this paper the theoretical analysis is reviewed. Then, computer simulation, including 
its algorithm, calculation and generated maps, is presented. The theoretical 
analysis is validated by studying these maps.

\vspace{1ex}
\noindent
{\bf keywords.} aperture synthesis radio telescope -- instrumentation: interferometers 
	-- high order harmonic interferometer -- computer simulation
		\end{abstract}
\bigskip			
	\end{@twocolumnfalse}
]

\section{Introduction}
In an aperture synthesis radio telescope (ASRT), a correlator is used for each baseline 
to generate a Fourier component of the radio map, that is, the distribution of power density  (brightness) of the field of view (FOV) in the sky under observation. 
The map is reconstructed by the Fourier transform (FT) technique.

The resolvability of an ASRT depends mainly on the maxi\-mum baseline. Therefore, 
conventionally the increase of resol\-vability is achieved by extending the maximum 
baseline. A new technique was proposed by \citet{wu1996}, in which the increase of 
resolvability was achieved by using a new type of interfero\-meter, called the 
\emph{High Order Harmonic Interferometer} (HOHI), without extending the 
maximum baseline. 

This paper is devoted particularly to the HOHI of second order. Its theory is 
reviewed first. For completeness, some mathema\-tical formulas and a figure in 
\citet{wu1996} are reproduced, mutatis mutandis. 
Then, computer simulation is presented, including its algorithm, 
calculation and generated maps. These maps are stu\-died in detail and the theory 
of HOHI is validated. Some practical issues in the implementation of HOHI are discussed.

\section{Theory of HOHI}

\begin{figure*}[ht]
\centering
\includegraphics[width=0.90\linewidth]{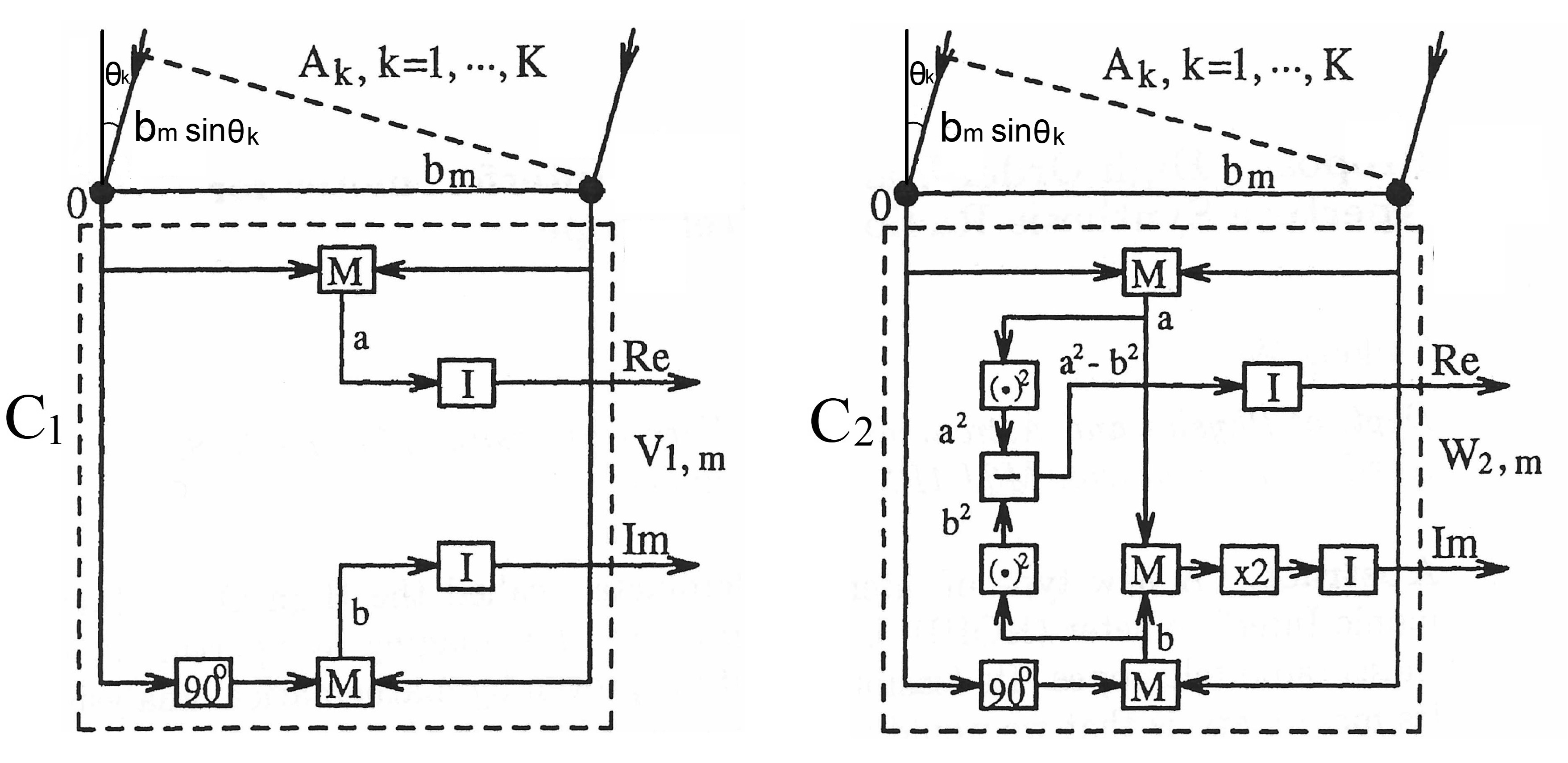} 
\caption{Correlators for the baseline $b_m$. \hspace{2pt} 
	$\mathrm{C}_1$:~Conventional (base).\,  $\mathrm{C}_2$:~Second order. 
	\protect\citep{wu1996} }
\label{fig:1}
\end{figure*}
 
In the following, one-dimensional (1-D) notation is used for simp\-licity and clarity. 
As depicted in Fig.~\ref{fig:1} , $\mathrm{C}_1$ and $\mathrm{C}_2$ are respectively
the conventional (base or first harmonic) correlator and the second harmonic correlator,
where  $b_m$ ($m =1,...,M$) is a baseline, $b_M$ being its maximum; 
the block M is multiplier, I is integrator (time averager), and the  $90^\mathrm{o}$ 
block is phase shifter. 
    
Suppose in the FOV there are $K$ point sources ha\-ving amplitudes $A_k$ and 
random phases $\phi_k,\, k = 1,...,K$. The radiations of wavelength $\lambda$ 
from a single point source $A_k$ arrive at the angle $\theta_k$ with respect to the 
normal to the baseline. Then, the phase difference is $2\pi u_m x_k$ 
between the two signals in the left and right branches for the baseline $b_m$, where 
$u_m = b_m/\lambda$ (baseline measured in wavelengths), and the coordinate 
$ x_k = \sin\theta_k$ (instead of $\theta_k$) is used to specify the angular position 
of point source $k$.

The output from $\mathrm{C}_1$  is 
\begin{equation} \label{eq:1}
V_{1,m} = W_{1,m} = \frac{1}{2} \sum_{k=1}^K A_k^2 \mathrm{e}^{-i2\pi u_m x_k}\; .
\end{equation}
Here $i=\sqrt{-1}$. $V_{1,m} $ is the (complex) visibility output from the conventional interferometer (the spatial frequency being $1 \times u_m =u_m$). 
The output from $\mathrm{C}_2$  is 
\begin{equation}\label{eq:2}
 W_{2,m} = \frac{1}{4} \left\{  \sum_{k=1}^K A_k^4  \mathrm{e}^{-i2\pi 2u_m x_k} + 
 	4\sum_{p=1}^K  \sum_{q>p}^K A_p^2 A_q^2 \mathrm{e}^{-i2\pi u_m (x_p+x_q)}  \right\}\; ,
\end{equation}
which consists of the desirable second harmonics and unwanted cross terms.
The latter can be eliminated by operations as shown in the following:
\begin{equation}\label{eq:3}
V_{2,m} = 2 V_{1,m}^2 - W_{2,m} = 
	\frac{1}{4} \sum_{k=1}^K A_k^4 \mathrm{e}^{-i2\pi 	2u_m x_k}\; .
\end{equation}
$ V_{2,m}$ is the visibility output for the baseline $b_m$ from the HOHI of second order. 
Its spatial frequency is $2  u_m$, that is to say, every baseline is virtually increased by 
a factor of 2. In particular, let $m=M$, the maximum baseline, $b_M$, is virtually doubled. 
Therefore, the resolvability of the telescope is also doubled.

\section{Computer simulation of HOHI}

The purpose of computer simulation is to generate 1-D maps using visibilities 
$V_{1,m}, m =1,...,M$ from the conventional inter\-ferometer (called \emph{base}) 
and $V_{2,m}, m =1,...,M$ from the HOHI of second order (called \emph{2nd-order}). 
A number of cases are studied. In each case the base map is compared with 
the 2nd-order map.

Our ``virtual'' ASRT has an east-west antenna array with the following parameters:
wavelength $\lambda = 0.21$ meter, 
the number of baselines (channels) $M = 144$, 
the baseline increment $d = 4.286$ meters, and 
the maximum baseline $b_M = 617.184$ meters.
They are taken from the synthesis telescope at Dominion Radio Astrophysical 
Observatory (DRAO, \citealt{landecker2000}) . 
 
The visibility used to reconstruct maps has $L=2 M + 1 = 144\times 2+1=289$ data points, 
where the factor $2$ accounts for the Hermitian property of visibility 
(each $V_{1,m}$ or $V_{2,m}$ from the interferometers has its complex conjugate); 
the additional $1$ accounts for the zero-baseline component. 
If this visibility is used for the FT without zero-padding, the pixel size $\Delta x$ of 
the map would be $\lambda / d / L = 0.21/4.286/289 = 1.6954\times 10^{-4}$.
This pixel size is called \emph{unit} (pure number). 
In order to increase the accuracy of measurement, zero-padding to the visibility
must be used to interpolate the gene\-rated maps, that is, to reduce the pixel sizes of maps.
These reduced pixel sizes will be expressed in terms of unit.

\begin{figure}[ht]
\centering
\includegraphics[width=0.72\linewidth]{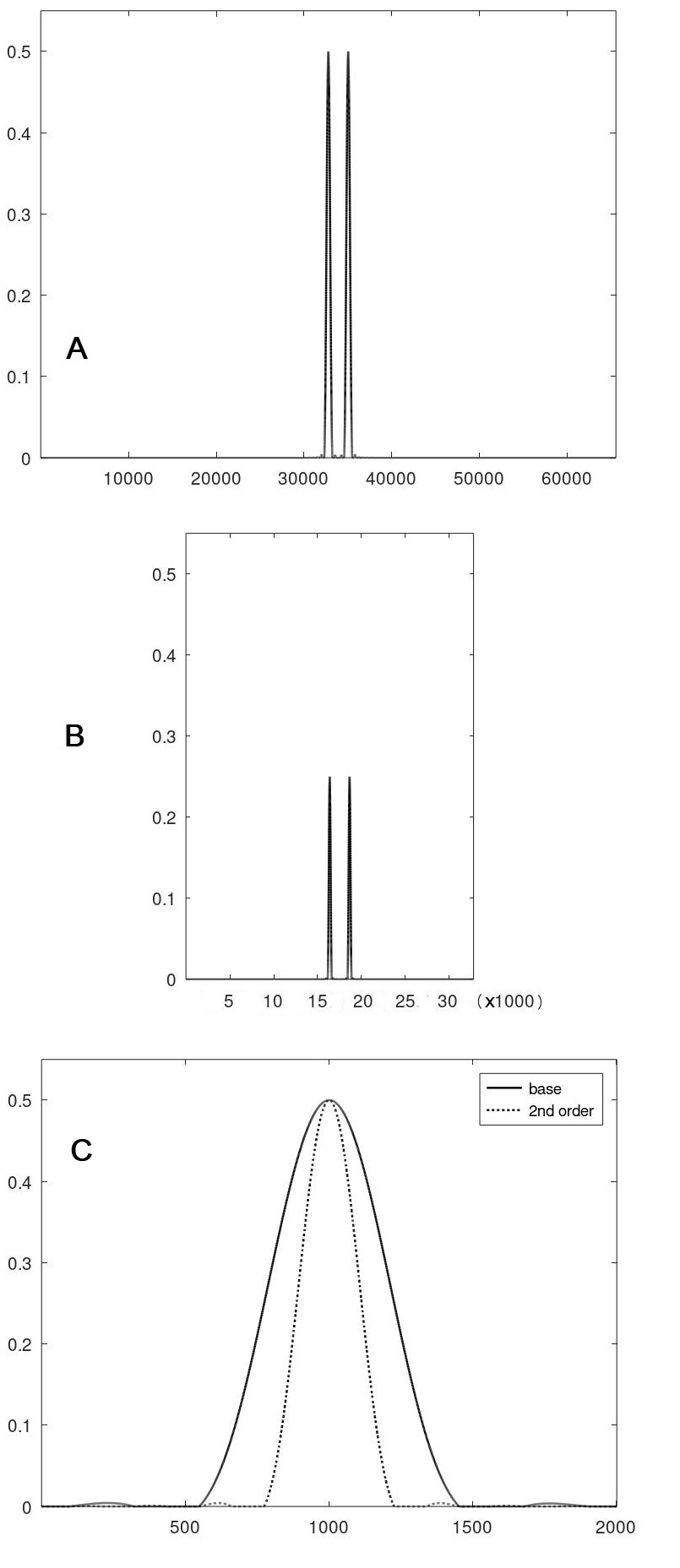} 
\caption{Case 1. Maps of two well-resolved point sources. 
	{\bf{A}}. Base. FT length=65536. 
	{\bf{B}}. 2nd-order. FT length=32768. 
	{\bf{C}}. Left peaks in the central parts of {\bf{A}} (solid line) 
		and {\bf{B}} (dotted line). Length=2001.}
\label{fig:2}
\end{figure}
  
Simulation is carried out using the Octave language \citep{eaton2022}.
Four cases are studied. 
They are: {\bf{1}}. Two point sources are located far apart so that they are well resolved 
in both the base and 2nd-order maps. {\bf{2}}. Two point sources are located closely 
such that they are not resolved in the base map but resolved in the 2nd-order map. 
{\bf{3}}. Two point sources are located so closely that they are not resolved 
in either the base or 2nd-order maps. {\bf{4}}. An extended source.

\subsection{Case 1. Two point sources,  well resolved}
 
The purposes of this study are to calculate the Full Widths at Half Maximum (FWHMs) in 
the base and 2nd-order maps, and to compare them. 

Two point sources have the same amplitude of 1.0 and are located at 0 unit 
(at the center of the FOV, $x=0,\, \theta=0$ radian) and 10.0 units, respectively. 
This sufficiently large distance between the two sources ensures that they are well resolved in the reconstructed maps. 

First of all, we determine the necessary FT length, $L_1$, for the base map  to ensure 
the accuracy of measurement. 
The measurement of FWHM we implement using Octave has the maximum error of two pixels.
The required measurement accuracy is $0.01$ unit. Therefore, in the worst case we have
$2 \times 289/L_1 = 0.01$, and $L_1 = 2 \times 289 / 0.01 = 57800$. 
For the most efficient Fast Fourier Transform (FFT), this number is run up to a power of 2, 
and we get $L_1 = 65536 = 2^{16}$. Correspondingly, the accuracy is 
$2 \times 289/65536 = 0.0088$ that is better than $0.01$.
On the other hand, in comparison with the base map, the 2nd-order map has effectively 
the doubled maximum baseline and hence a halved pixel size. As a result, the required 
FT length is $L_2 = L_1/2 = 2^{15} = 32768$.  

For generating the base map, the visibilities $V_{1,m}, m=1,...M$  calculated using 
Eq.~\ref{eq:1} and their complex conjugates are used as the positive and negative parts
of the FT components, respectively. The zero-baseline visibility (total power) 
is unavai\-lable, but we know that it will be greater than the absolute values of
other visibilities. Therefore, the maximum of the absolute values of 
$V_{1,m}, m=1,...M$ will be taken as the zero-baseline visibility, that is, the zero-frequency
component. By the way, this component will determine the background of the 
reconstructed map, but will not affect the resolvability. 

The resultant $2M+1$ point visibility is tapered by the Hanning window, and then 
zero-padded to obtain the required FFT length. Finally, the FFT is performed to
generate the map.

Shown in Fig.~\ref{fig:2} \,are three maps.  
{\bf A}. Base map. It is reconstructed using the base 
visibility as described in the preceding paragraphs. The FT length is $65536$. 
The map center is at $32769$.
 {\bf B}. 2nd-order map. It is gene\-rated in the same way as generating {\bf A}, except
 that the visibilities $V_{2,m}, m=1,...M$  calculated using Eq.~\ref{eq:3} are used 
and the FFT length is $32768$. The map center is at $16385$. 
These two maps have the same pixel size  $0.0044$ unit.

In the HOHI of second order all the baselines, including the maxi\-mum baseline, 
are stretched by a factor of 2. Meanwhile, the baseline increment (sampling interval)
is also increased by a factor of 2. Therefore, the resolvability is doubled while the 
width of synthesized field (WSF) is halved. Specifically, for the base map 
$d = 4.286$ meters, and the $\mathrm{WSF} = \lambda / d = 0.21/4.286 = 0.0490$.
In contrast, for the 2nd-order map
the $\mathrm{WSF} =\lambda / (2d) = 0.0245$. This is why the horizontal axis
length of map~{\bf B} is a half of that of map~$\bf A$.

As expected, the images of the point sources have peak va\-lues $0.5$ and $0.25$
in maps~{\bf A} and~{\bf B}, respectively.
Furthermore, the images in map~{\bf B} have narrower peaks compared
with the images in map~$\bf A$. This means that the resolvability in map~{\bf B} 
is better. This improvement of resolvability can be seen clearly in map~{\bf C}.

Shown in map~{\bf C} are the left peaks in the central parts of 
maps~{\bf A}  (solid line) and {\bf B} (dotted line). Note that 
the latter has been scaled for pixel values so its maximum is equal to 
that of the former for the ease of visual comparison and calculation.
Quantitatively, the FWHMs of the 2 peaks are $451$ and  $225$ pixels, respectively.
Their ratio is $451/225 = 2.00$, that is, the resol\-vability of the 
2nd-order map is improved by a factor of 2.

\subsection{Case 2. Two point sources, located closely}   

The purpose of this study is to visually show that in the critical case the resolvability
in the base map and that in the 2nd-order map are dramatically different.

\begin{figure}[t]
\centering
\includegraphics[width=0.72\linewidth]{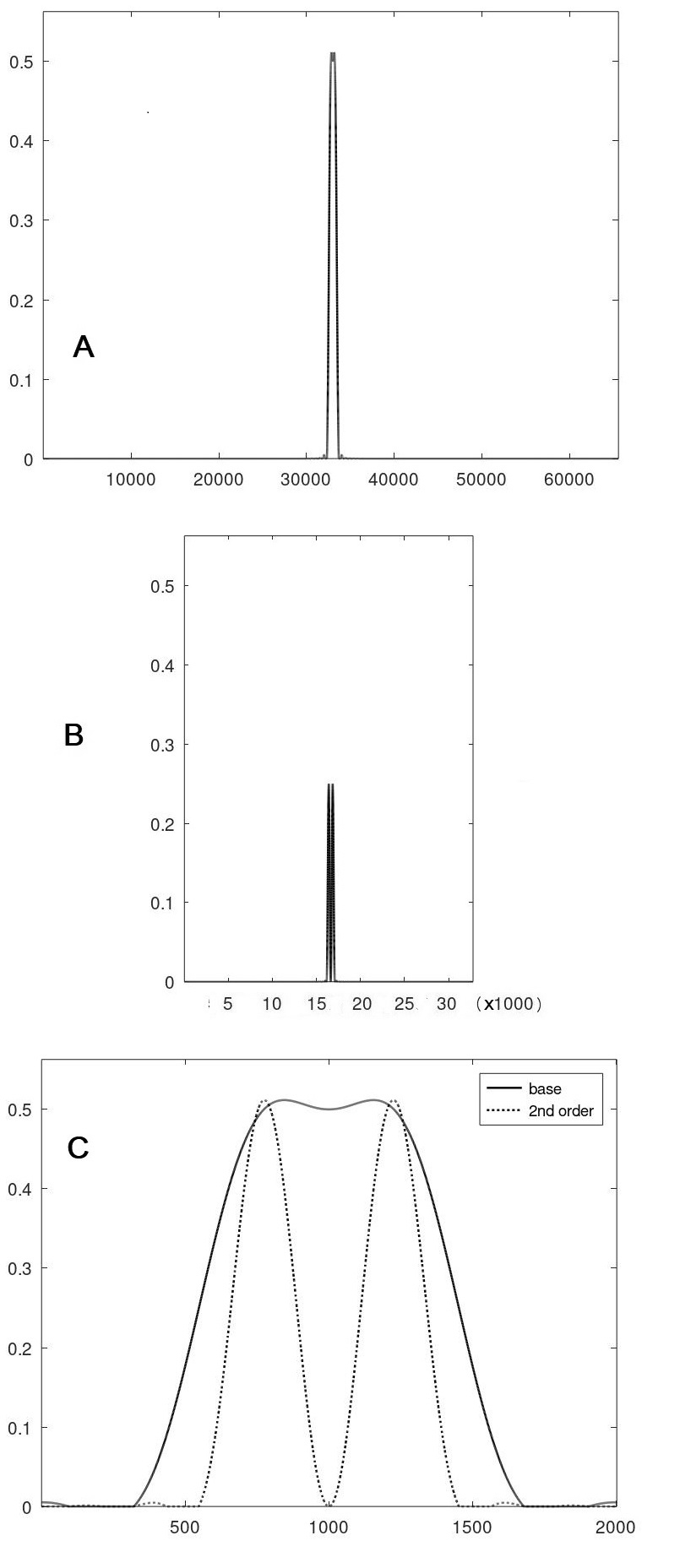} 
\caption{Case 2. Maps of two closely-located point sources. 
	{\bf{A}}. Base. FT length=65536.  
	{\bf{B}}. 2nd-order. FT length=32768. 
	{\bf{C}}. The central parts of {\bf{A}} (solid line) and {\bf{B}} (dotted line).
	 	Length=2001.}
\label{fig:3}
\end{figure}

Two point sources have the same amplitude of 1.0 and are located at 0 unit 
and 2.0 units, respectively. Three maps, as shown in Fig.~\ref{fig:3}, are generated 
in the same way as in Case~1. It can be seen in maps~{\bf A} and~{\bf B}
that the 2 peaks are well resolved in the 2nd-order map but not resolved in the base map.
This difference can be seen easily in map~{\bf C}. 
 
\subsection{Case 3. Two point sources, not resolved}   

\begin{figure}[t]
\centering
\includegraphics[width=0.72\linewidth]{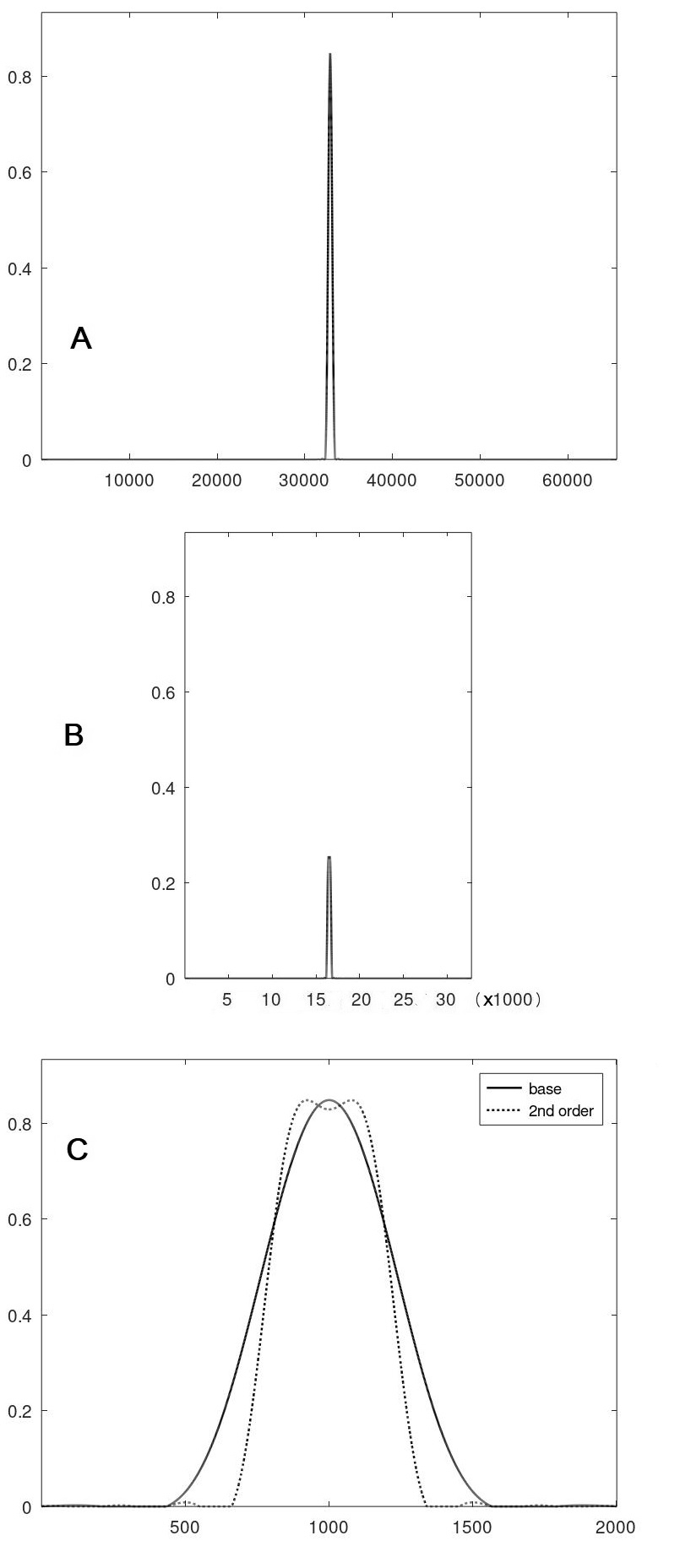} 
\caption{Case 3. Maps of two not-resolved point sources. 
	{\bf{A}}. Base. FT length=65536.  
	{\bf{B}}. 2nd-order. FT length=32768. 
	{\bf{C}}. The central parts of {\bf{A}} (solid line) and {\bf{B}} (dotted line).
	 	 Length=2001.}
\label{fig:4}
\end{figure}
    
The purpose of this study is to visually show the case in which two point sources are 
so closely located that they are not resolved in either the base or 2nd-order 
maps. 
 
Two point sources have the same amplitude of 1.0 and are located at 0 unit 
and 1.0 unit, respectively. Three maps, as shown in Fig.~\ref{fig:4}, are generated 
in the same way as in Case~1.
It can be seen in map~{\bf C} that the 2 peaks of the base map are completely 
merged to one peak (solid line). In contrast, the 2 peaks of the 2nd-order map are not completely merged  (dotted line) , and the top center is convex downward, 
which is an indication that 2 distinct peaks might exist.

\subsection{Case 4. Extended source}   

\begin{figure}[t]
\centering
\includegraphics[width=0.72\linewidth]{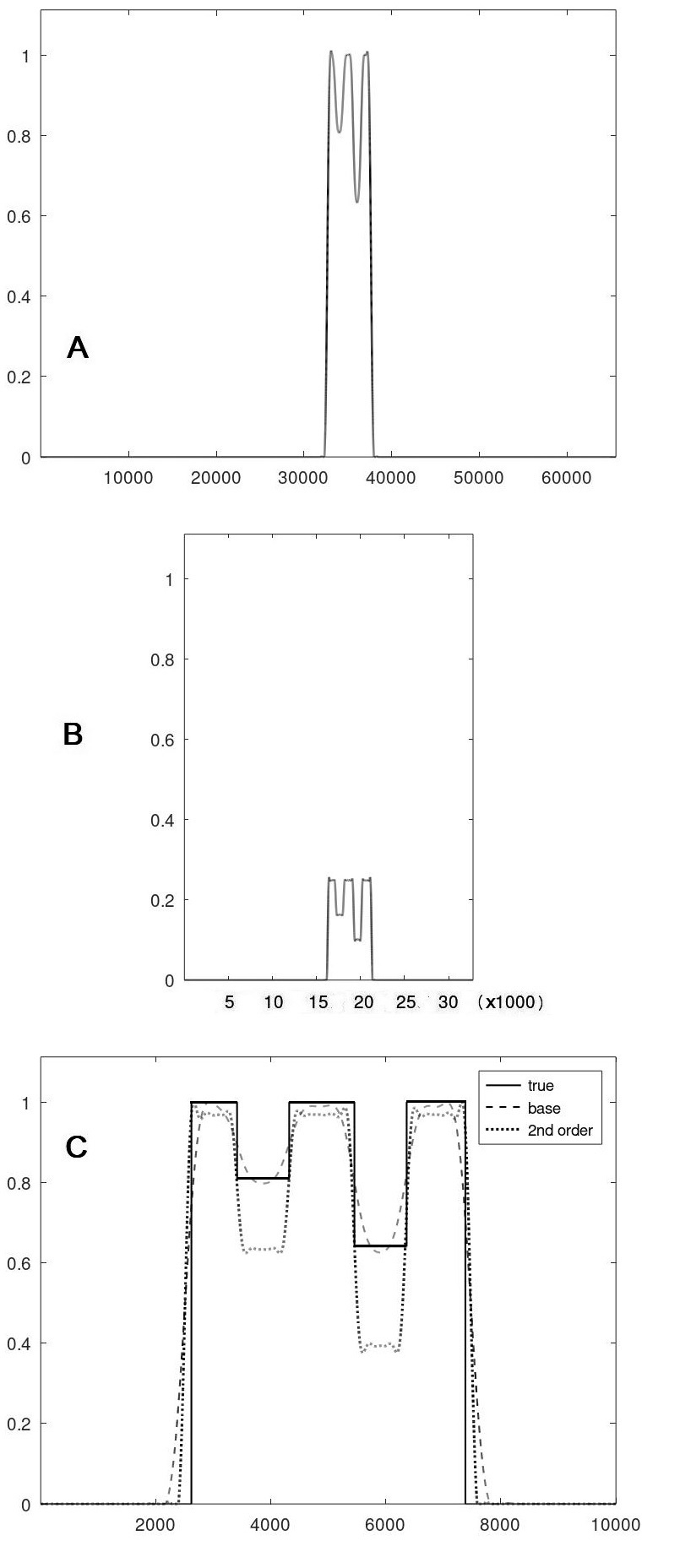} 
\caption{Case 4. Maps of an extended source. 
	{\bf{A}}. Base. FT length=65536.  
	{\bf{B}}. 2nd-order. FT length=32768. 
	{\bf{C}}. The central parts of the true image (solid line), {\bf{A}} (dashed line) and
		{\bf{B}} (dotted line). Length=10001.}
\label{fig:5}
\end{figure}
        
The purpose of this study is to compare, with the ``true image'' as a reference,  
the images of an extended source in the base and 2nd-order maps. 
The extended source is a plateau having a flat top with 2 notches 
(see Fig.~\ref{fig:5}.{\bf C}, solid line). 
The point-source model is used for the extended source, that is, it is represented with a number of points.

Three maps, as shown in Fig.~\ref{fig:5}, are 
generated in the same way as in Case~1. The central parts are shown in 
map~{\bf C}. Shown in this map additionally is the central part of the true image.

It can be seen in the 2nd-order map~{\bf B} that the image has sharp edges.
In contrast, the image's edges in the base map~{\bf A} are unclear. This difference
can be seen more easily in map~{\bf C} with the true image as the reference.

\section{Discussion}

In this section we discuss some practical issues in the implementation of HOHI.

\subsection{Generation of the second order harmonics}

As shown in Fig.~\ref{fig:1}, the conventional correlator $\mathrm{C}_1$ 
and second order correlator  $\mathrm{C}_2$ 
share the same preprocessed signals as their inputs. With the existing ASRT, for each baseline we need only to build $\mathrm{C}_2$ 
according to the formula $(a+ib)^2=(a^2-b^2)+i2ab$. 
The visibility $V_{2,m}$ is gene\-rated according to Eq.~\ref{eq:3}.

\subsection{Baselines and antenna array}

Usually, the increment $d$ is chosen critically in designing an antenna array. 
The WSF determined by $\lambda/d$ and the FOV determined by antenna parameters 
have been carefully chosen in order to eliminate aliasing caused by strong 
radio sources inside the FOV but outside the WSF. 
The reduction of WSF may cause the aliasing problem.

In terms of signal processing, $d$ is the sampling interval. Doub\-ling it 
may cause the problem of undersampling, and hence alia\-sing,
depending on the lowpass filter effect of the antenna beam pattern.

In order to avoid this aliasing problem, the increment after being stretched 
must be kept as designed. Specifically, the increment of the 
\emph{real baseline} must be $\frac{1}{2} d$ that will be stretched to $d$. 

\begin{figure}[t]
\centering
\includegraphics[width=0.95\linewidth]{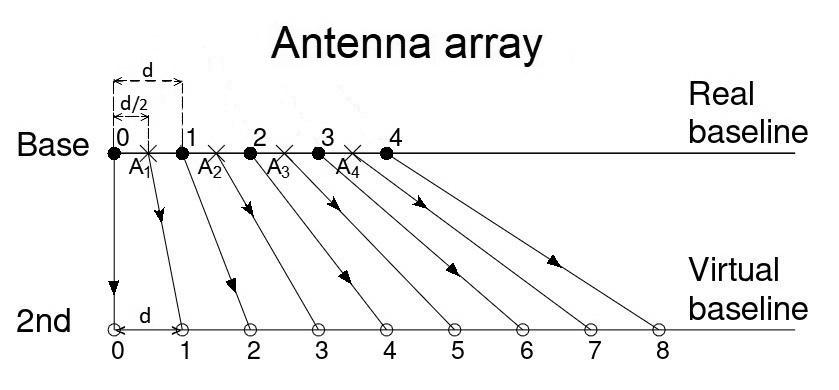} 
\caption{Antenna array and baselines. Base axis: Dots represent the exis\-ting antenna
	positions; crosses represent the added antenna positions. 
	\,2nd axis: Circles represent all the virtual positions of antennas.} 
\label{fig:6}
\end{figure}

Let us look into the simplest east-west antenna array that consists of two antennas, 
one being fixed at the west end (antenna 0) while the other being movable, as shown schematically on the ``Base'' axis in Fig.~\ref{fig:6}. 
 The movable antenna takes 4 positions, $1, 2, 3$ and $4$, which provides the 4 
 ``real baselines'' 
 $\overline{0-1}, \overline{0-2}, \overline{0-3}$ and $\overline{0-4}$ with the 
 increment $d$. The corresponding 4 ``virtual baselines'' are
  $\overline{0-2}, \overline{0-4}, \overline{0-6}$ and $\overline{0-8}$ with the 
 increment $2d$,  as shown on the ``2nd'' axis in Fig.~\ref{fig:6}. 
 For the increment after being stretched to be $d$,
 the movable antenna takes $4$ more positions,  
$\mathrm{A}_1, \mathrm{A}_2, \mathrm{A}_3$ and $\mathrm{A}_4$, 
represented by the $4$ crosses at the midpoints of 
$\overline{0-1}, \overline{1-2}, \overline{2-3}$ and $\overline{3-4}$, respectively. 
As a result, 4 additional virtual baselines are gene\-rated.
They are $\overline{0-1}, \overline{0-3}, \overline{0-5}$ and $\overline{0-7}$ 
on the 2nd axis. The whole set of 8 virtual baselines, 
 $\overline{0-1}, \overline{0-2}, \dots, \overline{0-8}$, provides 
the desirable increment $d$.

For this simplest antenna array, doubling the number of
baselines means doubling the number of positions the movable antenna
takes, that is to say, the observation time will be doubled. This is the 
price we must pay for the improved resolvability as if the maximum baseline
of array were doubled. For other antenna array configurations, this proportionality
may not hold. The required number of positions the movable antennas
take, and hence the observation time, may be increased by a factor of
less than $2$ by utilizing ``unused'' baselines in the antenna spacings.
On the other hand, in the general case, some required additional positions 
of the movable antennas may not be realizable because of physical constraints.
As a consequence, some required virtual baselines may be missing. 

\subsection{Test at DRAO}

The telescope of DRAO is suitable for the test of HOHI. Its li\-near array consists of 
$7$ east-west antennas, of which $2$ are fixed respectively at the $2$ ends of the array, 
and the other $5$ are movable along the rail. As a matter of fact, some basic para\-meters
used in our computer simulation are taken from this antenna array, namely, 
$\lambda = 0.21$ meter, $d = 4.286$ meters, and  $b_M = 617.184$ meters.

Because of the constraint imposed by the terrain surroun\-ding  DRAO, it is very difficult, 
if not impossible, to extend the antenna array. Therefore, it makes sense to 
improve the resol\-vability  of the telescope utilizing the technique of HOHI.  

As declared in \citet{landecker2019}
concerning the renewal of the telescope of DRAO,
it is an open testbed for Canadian innovation. We hope that our test will be
carried out at DRAO and make a significant contribution to the development 
of radio astronomy.

 \section{Conclusions}

Computer simulation has validated the theoretical analysis of the HOHI of second order. 
The resolvability of the telescope is improved by a factor of 2. Test may be carried out
utilizing facilities at DRAO.

\bibliography{HOHI_Simulation}

\end{document}